\documentstyle[aps,tighten,preprint,psfig]{revtex}
\newcommand{\beq}    {\begin{equation}}
\newcommand{\eeq}    {\end{equation}}
\newcommand{\beqarr} {\begin{eqnarray}}
\newcommand{\eeqarr} {\end{eqnarray}}
\newcommand{\barr}   {\begin{array}}
\newcommand{\earr}   {\end{array}}

\newcommand{\lsim}{\mathrel{\mathop{\kern 0pt \rlap
  {\raise.2ex\hbox{$<$}}}
  \lower.9ex\hbox{\kern-.190em $\sim$}}}
\newcommand{\gsim}{\mathrel{\mathop{\kern 0pt \rlap
  {\raise.2ex\hbox{$>$}}}
  \lower.9ex\hbox{\kern-.190em $\sim$}}}

\begin{document}

\preprint{
\begin{tabular}{r}
DFTT 32/2001 \\
\end{tabular}
}

\title{Size of the neutralino--nucleon cross--section in the light of
  a new determination of the pion--nucleon sigma term}

\author{\bf 
A. Bottino$^{\mbox{a}}$,
F. Donato$^{\mbox{b}}$, 
N. Fornengo$^{\mbox{a}}$,
S. Scopel$^{\mbox{a}}$
\footnote{E--mail: bottino@to.infn.it, donato@lapp.in2p3.fr, 
fornengo@to.infn.it, \\
\phantom{E--mail:~~~} scopel@to.infn.it}
\vspace{6mm}
}

\address{
\begin{tabular}{c}
$^{\mbox{a}}$
Dipartimento di Fisica Teorica, Universit\`a di Torino \\
and INFN, Sez. di Torino, Via P. Giuria 1, I--10125 Torino, Italy\\
$^{\mbox{b}}$
Laboratoire de Physique  Th\'eorique LAPTH \\
Annecy--le--Vieux, 74941, France \\
\end{tabular}
}


\maketitle

\begin{abstract}
We discuss the implications of a new determination of the
pion--nucleon sigma term for the evaluation of the neutralino-nucleon
cross--section, and, in turn, for detection rates for relic
neutralinos in WIMP direct experiments and in  some of the indirect
searches. 
It is shown that  the new range for the pion--nucleon sigma
term, taken  at its face value, favours values of the 
 neutralino-nucleon
cross--sections which are sizeably larger than  some  of the
current estimates. Implications for neutralino cosmological properties
are derived. 
\end{abstract}

\section{Introduction}

The evaluation of detection rates for relic neutralinos in WIMP direct
experiments and in measurements of signals due to possible 
neutralino--neutralino annihilation in Earth and Sun \cite{bf}
requires a reliable calculation of the neutralino--nucleon cross--section. 
 
In Ref. \cite{bdfs} we stressed that sizeable uncertainties in the  
neutralino--nucleon cross--section are due to the estimate of the
hadronic matrix elements, which describe the quark density contents in
the nucleon; this point was also remarked in 
Refs. \cite{don,jkg}.  In \cite{bdfs}  we quantified the range of these
uncertainties and showed that, within their ranges, the couplings of
neutralinos to nucleon may be sizeably stronger than some of those 
currently employed in the literature. Subsequently, an analysis of 
this problem was also undertaken in Refs. \cite{ellis,arnowitt,nath,fmw}. 

One of the crucial ingredients 
in the calculation of the hadronic matrix elements is the so--called
pion--nucleon sigma term. Its   numerical derivation from experimental 
data of pion--nucleon scattering is rather involved, and thus the origin
of  considerable uncertainties. Recent experimental results in
pion--nucleon scattering have now prompted a new determination of the 
pion--nucleon sigma term \cite{pavan}. In the present paper, we
consider the implications 
of these new inputs  for the neutralino--nucleon cross--section; in
particular, we show that the new data reinforce our previous
conclusions of Ref. \cite{bdfs}. As compared to Ref. \cite{bdfs}, the
present paper, apart from the use of the new evaluation for the sigma term, 
 contains also some other relevant updatings, mainly: i)
radiatively corrected Higgs--quark couplings \cite{qcd_corr} and ii)
recent experimental bounds on Higgs masses and supersymmetric
parameters from LEP2 \cite{lep} and CDF \cite{cdf}.

\section{Neutralino--nucleon elastic cross--section}

 The neutralino--nucleon scalar cross--section may be written as 

\beq
\sigma_{\rm scalar}^{(\rm nucleon)} = \frac {8 G_F^2} {\pi} M_Z^2 m_{\rm red}^2 
\left[\frac{F_h I_h}{m_h^2}+\frac{F_H I_H}{m_H^2}+
\frac{M_Z}{2} \sum_q <N|\bar{q}q|N>
\sum_i P_{\tilde{q}_i} ( A_{\tilde{q}_i}^2- B_{\tilde{q}_i}^2)
\right]^2.  
\label{eq:sigma}
\eeq

\noindent
The first two terms inside the brackets refer to the diagrams with 
exchanges of the two CP--even  neutral Higgs bosons, $h$ and $H$, 
in the t--channel (the exchange diagram of the CP--odd one, $A$, 
is strongly kinematically suppressed and then omitted
here) \cite{barbieri} and the third term refers 
to the graphs with squark--exchanges in the s-- and u--channels \cite{griest}.
The mass $m_{red}$ is the neutralino--nucleon
reduced mass. Since, for simplicity, in the present paper we
explicitly discuss only the Higgs--mediated terms, we do not report 
the expressions for the squark propagator
$P_{\tilde{q}_i}$ and for the couplings $A_{\tilde{q}_i}$, $B_{\tilde{q}_i}$,
which may be found in Ref.\cite{noi1234}. However, 
 in the numerical results reported in this paper also 
the squark--exchange terms are included. 
The quark matrix elements $<N|\bar{q}q|N>$ are meant over the nucleonic
state. 

The quantities $F_{h,H}$ and 
$I_{h,H}$ are defined as follows 
\beqarr
F_h &=& (-a_1 \sin \theta_W+a_2 \cos \theta_W) 
          (a_3 \sin \alpha + a_4 \cos \alpha) 
           \nonumber \\
           F_H &=& (-a_1 \sin \theta_W+a_2 \cos \theta_W) 
           (a_3 \cos \alpha - a_4 \sin \alpha) \nonumber \\
           I_{h,H}&=&\sum_q k_q^{h,H} m_q \langle N|\bar{q} q |N\rangle, 
\label{eq:effe}
\eeqarr
\noindent 
where the $a_i$'s are the coefficients in the definition of the 
 neutralino as the lowest--mass linear 
superposition of photino ($\tilde \gamma$),
zino ($\tilde Z$) and the two higgsino states
($\tilde H_1^{\circ}$, $\tilde H_2^{\circ}$)
\begin{equation}
\chi \equiv a_1 \tilde \gamma + a_2 \tilde Z + a_3 \tilde H_1^{\circ}  
+ a_4 \tilde H_2^{\circ} \, . 
\label{eq:neu}
\end{equation}

\noindent
The coefficients $k_q^{h,H}$ for up--type and down--type quarks 
are given in Table 1, in terms of the angle 
$\beta$, defined as  $\tan\beta = <H^0_2>/<H^0_1>$ and the angle
$\alpha$, which 
rotates $H_1^{(0)}$ and $H_2^{(0)}$ into $h$ and $H$. In  Table I  
also included are the coefficients $k_q^{h,H}$ for the CP--odd Higgs boson
$A$, which are important for the evaluation of the neutralino relic
abundance. The entries include those radiative corrections which may be
sizeable at large $\tan\beta$. These corrections affect the couplings to
down--type quarks $k_{d-type}$, and are parametrized in terms of the quantity
$\epsilon\equiv 1/(1+\Delta)$, where $\Delta$ enters in the relationship
between the fermion running masses $m_d$ and the corresponding Yukawa couplings
$h_d$ \cite{qcd_corr}:
\begin{equation}
m_d=h_d <H_0^1>  (1+\Delta)
\label{eq:delta_d}
\end{equation}
These corrections take contributions mainly from gluino--squark,
chargino--squark and neutralino--stau loops \cite{qcd_corr}. 

\section{Evaluation of the quantities $ m_q \langle N|\bar{q} q |N \rangle $}

For the calculation of the quantities $ m_q \langle N|\bar{q} q |N \rangle $
it is useful to express them in terms of the pion--nucleon sigma term
\beq
\sigma_{\pi N} = \frac{1}{2} (m_u + m_d) <N|\bar uu + \bar dd|N>,
\eeq
\noindent 
of the quantity $\sigma_{0}$, 
related to the size of the SU(3) symmetry breaking, and 
 defined as
\beq
\sigma_{0}\equiv \frac{1}{2}(m_u+m_d) <N|\bar uu+\bar dd-2\bar ss|N>,
\eeq
\noindent and of the ratio $r=2 m_s/(m_u+m_d)$.

Assuming isospin invariance for quarks 
$u$ and $d$, 
the quantities $ m_q \langle N|\bar{q} q |N \rangle $ for light
quarks may be written as 

\beqarr
m_u <N|\bar uu|N>&\simeq& m_d <N|\bar dd|N> \simeq \frac{1}{2} \sigma_{\pi N}
\label{eq:condlight}\\
m_s <N|\bar ss|N>&\simeq& \frac{1}{2} r (\sigma_{\pi N} - \sigma_{0}).
\label{eq:condstrange}
\eeqarr 
For the heavy quarks $c$, $b$, $t$, use of the heavy quark
expansion\cite{svz} provides
\beqarr
m_c <N|\bar cc|N>&\simeq& m_b <N|\bar bb|N> \simeq m_t <N|\bar tt|N> 
\simeq \nonumber\\
&\simeq& \frac{2}{27} \left [ m_N - \sigma_{\pi N} + 
\frac{1}{2}r (\sigma_{\pi N} - \sigma_{0})\right],
\label{eq:condheavy}
\eeqarr
\noindent where $m_N$ is the nucleon mass. The quantities 
$I_{h,H}$ can then be rewritten as
\beq
I_{h,H} = k_{u{\rm -type}}^{h,H} g_u + k_{d{\rm -type}}^{h,H} g_d,
\label{eq:i}
\eeq
\noindent where
\beqarr
g_u &\simeq& m_l <N|\bar{l}l|N>+\;2\;  m_h <N|\bar{h}h|N> \nonumber\\
&\simeq& \frac {4} {27} (m_N + \frac {19}{8} \sigma_{\pi N} 
- \frac{1}{2}r (\sigma_{\pi N} - \sigma_{0})),
\eeqarr
\beqarr
  g_d &\simeq& m_l <N|\bar{l}l|N>+\; m_s <N|\bar{s}s|N>+\; m_h
  <N|\bar{h}h|N> \nonumber\\
&\simeq& \frac{2}{27} (m_N + \frac{23}{4} \sigma_{\pi N} 
+ \frac{25}{4} r  (\sigma_{\pi N} - \sigma_{0})); \label{eq:g}
\eeqarr

\noindent 
 $l$ stands for light quarks ($l=u,d$) and $h$ denotes the heavy
 ones ($h=c,b,t$). 
                 
We recall that from the values of 
$\sigma_{\pi N}$ and ${\sigma_{0}}$  one can derive 
the fractional strange--quark content of the nucleon $y$
\beq \label{eq:y}
y=2 \frac{<N|\bar ss|N>}
{<N|\bar uu+ \bar dd|N>},
\eeq

\noindent 
using the expression  
\beq
y=1-\frac{\sigma_{0}}{\sigma_{\pi N}}.
\label{eq:yy}
\eeq
                    
\subsection{Values for $\sigma_{\pi N}$, ${\sigma_{0}}$ and $r$}

The range of $\sigma_{\pi N}$ we used previously in Ref. \cite{bdfs}
is 

\beq
41\; {\rm MeV} \, \lsim \sigma_{\pi N}\lsim 57 \; {\rm MeV}.
\label{eq:k}
\eeq

\noindent
 This was derived from the 
pion--nucleon scattering amplitude, calculated at the so--called 
Cheng--Dashen point by Koch \cite{koch1}, and the evolution of the
nucleon scalar form factor,  as a function of the momentum
transfer from $t=2 m_{\pi}^2$ to $t=0$, evaluated in Ref. \cite{gls2}. 

We now consider the new determination of $\sigma_{\pi N}$ presented in
Ref. \cite{pavan}. The George Washington University/TRIUMF group,
using   up--dated pion--nucleon scattering data 
\cite{said} and  a new partial--wave and dispersion relation analysis  
program, has derived a  range for $\sigma_{\pi N}$ \cite{pavan}  

\beq
55 \; {\rm MeV} \lsim \sigma_{\pi N}\lsim 73 \; {\rm MeV}, 
\label{eq:p}
\eeq

\noindent
which turns out to be sizeably larger than the one of
Eq. (\ref{eq:k}). 
Values of the nucleon scalar form factor at the Cheng--Dashen point
higher than those of Ref. \cite{koch1} were also reported in 
Ref. \cite{olsson}. 

In the present  paper we consider the effect of employing the new range for 
$\sigma_{\pi N}$, given in Eq. (\ref{eq:p}), in the evaluation of 
the quantities $ m_q \langle N|\bar{q} q |N \rangle $'s. 

Here, ${\sigma_{0}}$ is taken in the range \cite{gl}

\beq
\sigma_{0}=30\div40 \;{\rm MeV}.
\label{eq:0}
\eeq

Also the mass ratio $r= 2 m_s/(m_u+m_d)$ may be affected by
significant uncertainties \cite{bdfs}. 
However, here, for consistency with some of the
previous determinations, we use the {\it default} value $r = 25$. 

It has to be noted  that combining together values of 
$\sigma_{\pi N}$ and ${\sigma_{0}}$ within their ranges in Eqs. 
(\ref{eq:p}) and (\ref{eq:0})  leads to rather large values for the 
fractional strange--quark content of the nucleon, as given by
Eq. (\ref{eq:yy}). This is a puzzle that urges further investigation  
in  hadron physics. In what follows, in the variations of the
quantities $\sigma_{\pi N}$ and ${\sigma_{0}}$
we impose the constraint that, anyway, $y \leq 0.5$.

\subsection{Values for $ m_q \langle N|\bar{q} q |N \rangle $, 
$g_u$ and $g_d$}

Inserting the numerical values of 
$\sigma_{\pi N}$, ${\sigma_{0}}$ and $r$ in the expressions given at
the beginning of this section, we finally obtain estimates for 
the quantities $ m_q \langle N|\bar{q} q |N \rangle $, 
$g_u$ and $g_d$. Since 
$m_s <N|\bar s s|N>$ is the most 
important term among the $m_q <N|\bar q q|N>$'s \cite{GGR}, unless 
$\tan \beta$ is very
small, the extremes
of the range of the neutralino--nucleon cross--section are provided by
the extremes of the range for $m_s <N|\bar s s|N>$. These, in turn, 
are given by: 
$(m_s <N|\bar ss|N>)^{min} \simeq \frac{1}{2} r ({\sigma_{\pi
    N}}^{min} - 
{\sigma_{0}}^{max})$ and 
$(m_s <N|\bar ss|N>)^{max} \simeq \frac{1}{4} r {\sigma_{\pi
    N}}^{max}$ (to satisfy the constraint $y \leq 0.5$). 
We call {\it Set b} the one with 
$m_s <N|\bar ss|N> = (m_s <N|\bar ss|N>)^{min}$ and {\it Set c} the one with 
$m_s <N|\bar ss|N> = (m_s <N|\bar ss|N>)^{max}$; we denote by {\it Set a}
the reference set of Ref. \cite{jkg}.  
The values of   the quantities $ m_q \langle N|\bar{q} q |N \rangle $, 
$g_u$ and $g_d$ 
are given in Table II.

We note that, with the new values of $\sigma_{\pi N}$, 
the coefficient $g_d$, which usually dominates in
the neutralino--nucleon cross--section, turns out to fall in the range 

\beq
266 \; {\rm MeV} \lsim  g_d \lsim 523 \; {\rm MeV}, 
\label{eq:gd}
\eeq

\noindent
to be compared with the reference value $g_d = 241$ MeV of {\it Set a}. Thus,
for most supersymmetric configurations, one expects 
$(\sigma_{scalar}^{(nucleon)})_{Set \;
  b}/(\sigma_{scalar}^{(nucleon)})_{Set \; a} 
\simeq 1.5$ and 
$(\sigma_{scalar}^{(nucleon)})_{Set
\;  c}/(\sigma_{scalar}^{(nucleon)})_{Set \; a} 
\simeq 6$.

\section{Results and conclusions}

The supersymmetric model employed in the present paper for the
calculation of the neutralino--nucleon cross--section and for the
neutralino relic abundance is the one
defined in Ref. \cite{probing,lat} and denoted there as effMSSM. We refer to
\cite{probing,lat} for details on the theoretical aspects and on the updated 
experimental bounds.

Inserting the values of Table II into Eq.(\ref{eq:sigma}), one obtains
the results displayed in Fig. 1. The two ratios in the cross--sections
are plotted as a function of $\xi \sigma_{scalar}^{(nucleon)}$, where
$\xi$ is a rescaling factor for the neutralino local density. 
 $\xi$ is
taken to be $\xi = {\rm min}\{1, \Omega_{\chi} h^2/(\Omega_m
  h^2)_{min}\}$, in order to have rescaling,
when $\Omega{\chi} h^2$ turns out to be less than $(\Omega_m h^2)_{min}$ (here
$(\Omega_m h^2)_{min}$ is set to the value 0.05). 

The use of $\xi \sigma_{scalar}^{(nucleon)}$ instead of simply 
$\sigma_{scalar}^{(nucleon)}$ allows one to better identify the range
of sensitivity in current WIMP direct experiments 
\cite{ge}, in Fig. 1. 
Taking into account astrophysical uncertainties \cite{belli}, this
range may 
established to  be 

\begin{equation}
4 \cdot 10^{-10} \; {\rm nbarn} \lsim \
\xi \sigma^{\rm (nucleon)}_{\rm scalar} \lsim 
 2 \cdot 10^{-8} \; {\rm nbarn}, 
\label{eq:exp}
\end{equation}

\noindent
for WIMP masses in the interval 
$40 \; {\rm GeV} \lsim  m_W \lsim 200 \;  {\rm GeV}$. In Fig. 1 we
notice that, in the sensitivity range of Eq.(\ref{eq:exp}), the
cross--section ratios are actually of the sizes obtained by our
previous estimate based on dominance of the term $g_d$, {\it i. e.}
$(\sigma_{scalar}^{(nucleon)})_{Set \;
  b}/(\sigma_{scalar}^{(nucleon)})_{Set \; a} 
\simeq 1.5$ and 
$(\sigma_{scalar}^{(nucleon)})_{Set
\;  c}/(\sigma_{scalar}^{(nucleon)})_{Set \; a} 
\simeq 6$. 

We turn now to the implications that our present analysis has for the
cosmological properties of relic neutralinos, as explored by present 
WIMP direct detection experiments. Fig. 2 provides the essential
information. The detection of relic neutralinos would be of great
relevance even if these particles constitute only a subdominant dark
matter population \cite{lat}; however,   it is obvious 
that the most attractive  case is when the neutralino 
relic abundance falls into the interval of cosmological interest: 
$0.05 \lsim \Omega_{m} h^2 \lsim 0.3$. Fig. 2 shows 
to what extent
the most interesting region: 
$0.05 \lsim \Omega_{m} h^2 \lsim 0.3$ and 
$4 \cdot 10^{-10} \; {\rm nbarn} \lsim \
\xi \sigma^{\rm (nucleon)}_{\rm scalar} \lsim 
 2 \cdot 10^{-8} \; {\rm nbarn}$, is covered by neutralino
 configurations, depending on the values of the input parameters of
 Table II.   
The up--right frontier of the scatter plots moves progressively 
upward, as we move from {\it Set a} of inputs to {\it Set b} and to 
{\it Set c}.

   The main results of the present paper may be summarized 
as follows:

\begin{itemize}

\item The new range of the pion--nucleon sigma term  favours 
values of the neutralino--nucleon cross--section which are 
sizeably larger 
(a factor of 1.5 to 6) than some of the current estimates. 
   However, a word of caution has to be said here:  
 the new derivation of the 
pion--nucleon sigma term implies a rather high value for the 
fractional  strange--quark content of the nucleon;  
this point requires  
further investigation in the framework of hadron physics.
Thus, for instance, the consistency between the present indication 
for higher values of the pion--nucleon 
sigma term and  the determination of the range of  $\sigma_{0}$ has 
 to be understood.

\item      

Uncertainties implied by the hadronic quantities 
 for the neutralino--nucleon cross--section 
are still quite sizeable; these have to be taken in due 
consideration in evaluations of the neutralino event rates.

\item 
The previous remark applies not only to event rates for 
WIMP direct detection, but also to the evaluation of the neutrino 
fluxes expected at neutrino telescopes, as a consequence of 
possible neutralino--neutralino annihilations inside the Earth 
and the Sun. In fact, these fluxes depend on the capture rate of 
neutralinos by the macroscopic bodies; in turn, this rate depends on 
the neutralino--nucleon cross--section.

\item A larger size of the neutralino--nucleon 
cross-section, as indicated by the new hadronic data, 
implies that present WIMP direct experiments (and some indirect 
measurements) explore larger sectors of the supersymmetric space where
the neutralino may be of cosmological interest.

\end{itemize}

\begin{table}
\caption{Values of the coefficients ${k_q}^{h,H}$  in 
Eq.(\ref{eq:effe}).
}
\begin{center}
\begin{tabular}{c|ccccc} 
~ & $h$ &~~ & $H$ &~~ & $A$ \\ \hline
$k_{u-type}$~~ & $\cos\alpha / \sin\beta$ &~~ & $\sin\alpha / \sin\beta$ &~~ & $1 /
\tan\beta$ \\
\\
$k_{d-type}$~~ & $-\sin\alpha / \cos\beta 
-\epsilon \cos (\alpha - \beta) \tan \beta$
&~~ & $\cos\alpha / \cos\beta-\epsilon \sin (\alpha - \beta) \tan
\beta$ &~~ 
& $\tan\beta (1+\epsilon)$
\end{tabular}
\end{center}
\end{table}

\begin{table}
\caption{Values of   the quantities $ m_q \langle N|\bar{q} q |N \rangle $, 
$g_u$ and $g_d$. 
{\it Set a} is the reference set given in
Ref.\protect\cite{jkg}. 
{\it Set b} and {\it Set c} are the sets corresponding to the minimal and maximal 
values of  $m_s <N|\bar s s|N>$, respectively. All quantities are in
units of MeV. 
}
\begin{center}
\begin{tabular}{c c c c c c}
 & $m_l<N|\bar{q}_l q_l|N>$ & $m_s<N|\bar{s}s|N>$ & $m_h<N|\bar{h}h|N>$ &
 $g_u$ & $g_d$ \\  \hline
{\it Set a} \cite{jkg}   & 27   & 131  & 56   & 139 & 214  \\
{\it Set b}              & 28   & 186  & 52   & 132 & 266  \\
{\it Set c}              & 37   & 456  & 30   &  97 & 523  \\
\end{tabular}
\end{center}
\end{table}

\newpage
\begin{figure}[t]
\hbox{
\psfig{figure=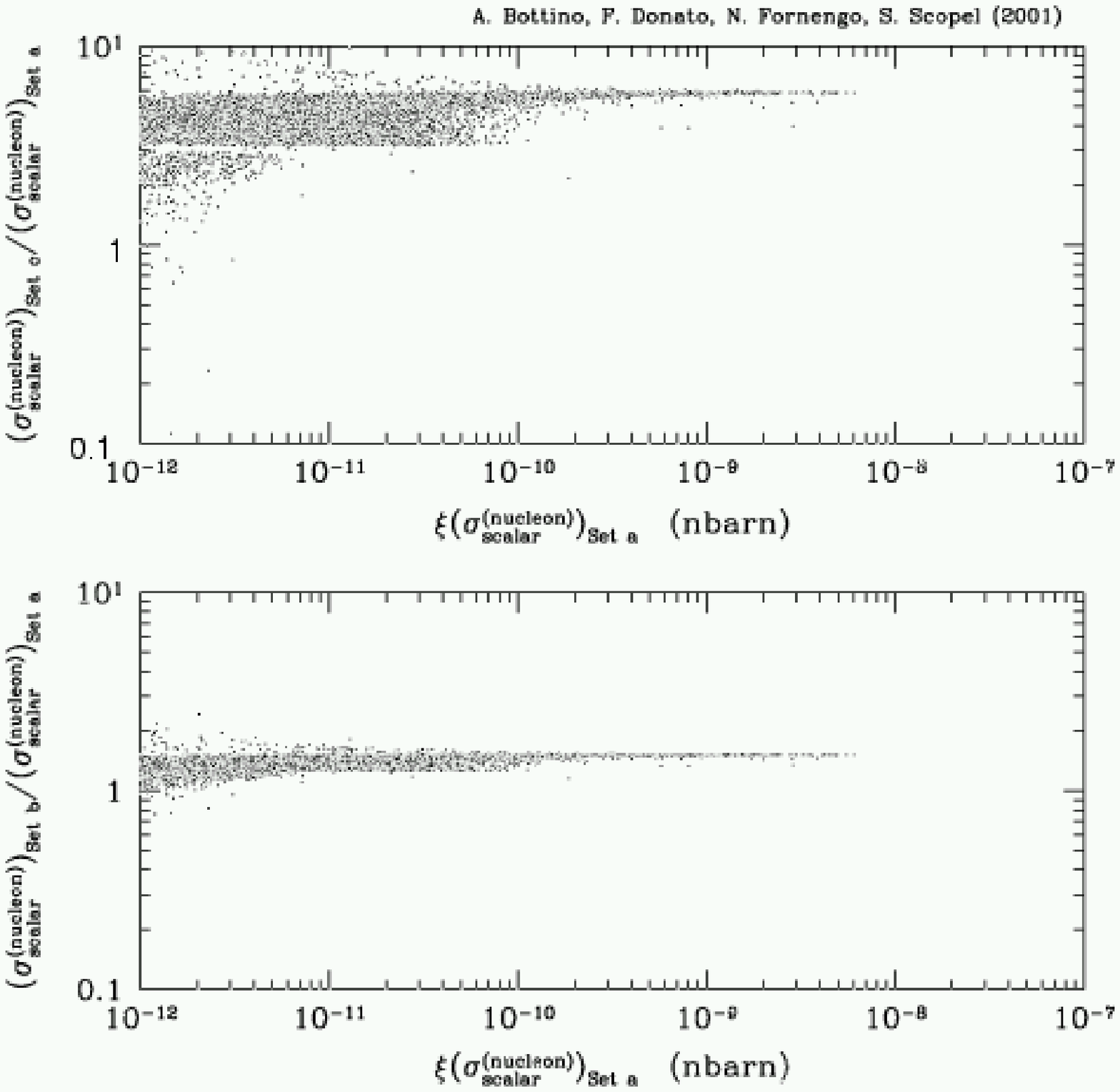,width=8.2in,bbllx=28bp,bblly=142bp,bburx=700bp,bbury=640bp,clip=}
}
{
FIG. 1.  Ratios of the neutralino--nucleon cross--sections, for values
of the quantities $ m_q \langle N|\bar{q} q |N \rangle $, 
$g_u$ and $g_d$, 
as given in Table II
}
\end{figure}

\newpage
\begin{figure}[t]
\hbox{
\psfig{figure=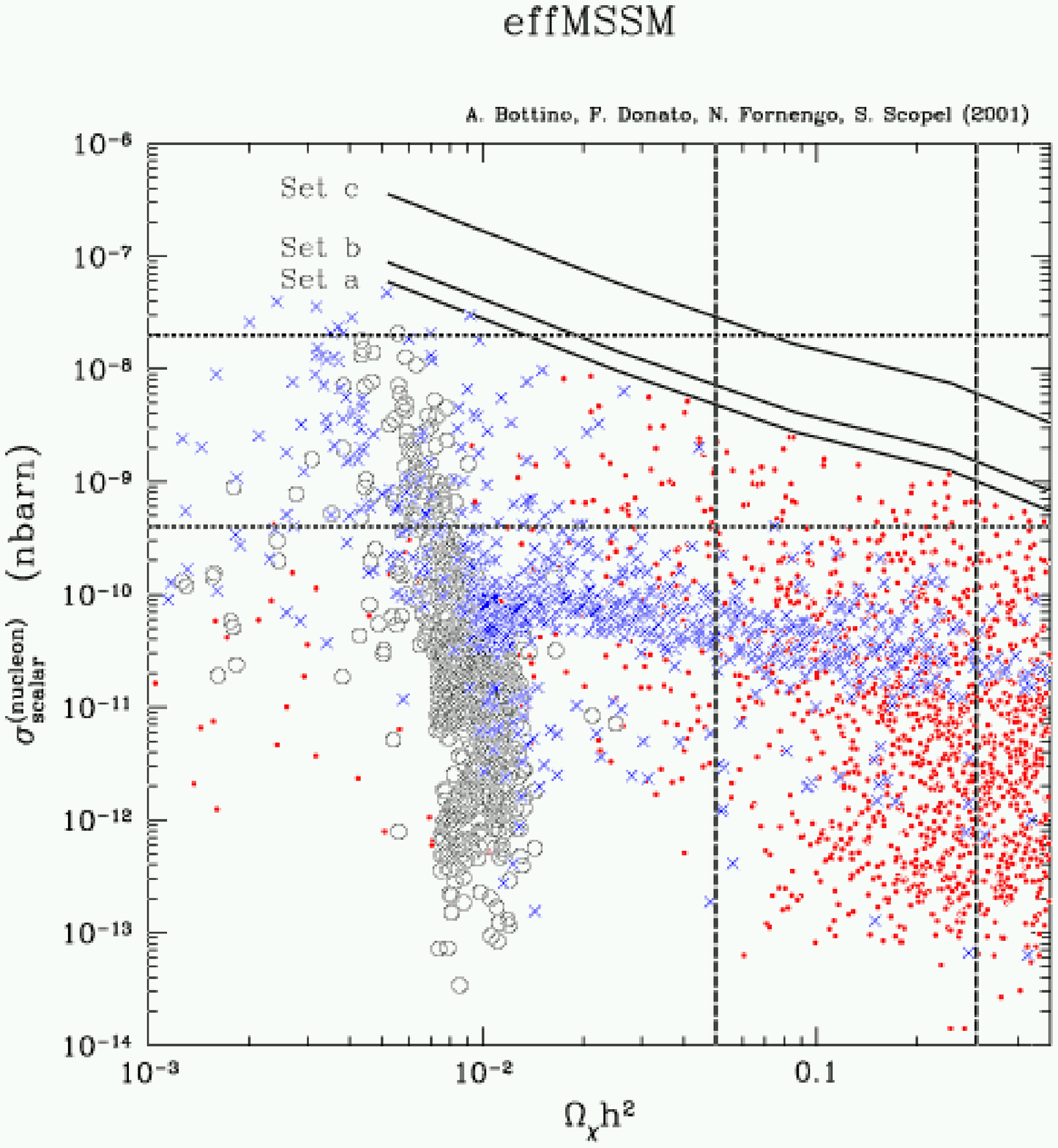,width=8.2in,bbllx=36bp,bblly=135bp,bburx=700bp,bbury=640bp,clip=}
}
{
FIG. 2.  
Scatter plot of $\sigma_{\rm scalar}^{(\rm nucleon)}$ versus 
$\Omega_{\chi} h^2$, when {\it Set a} is employed for the quantities given
in Table II. The curve denoted by {\it Set a} visualizes the up--right
frontier 
of the scatter plot. The curves denoted by {\it Set b} and by {\it Set c} give 
the locations of the frontiers of the scatters plots 
(not shown in this figure) obtained by using values of {\it Set b} and
{\it Set
c}, respectively. 
$m_{\chi}$ is taken in the range 
$40 \; {\rm GeV} \leq  m_{\chi} \leq 200 \;  {\rm GeV}$. 
 The two horizontal lines bracket the sensitivity region defined 
 by Eq. (\ref{eq:exp}), when the rescaling factor $\xi$ is set equal
 to one. 
 The two vertical lines denote the range 
$0.05 \leq \Omega_{m} h^2 \leq 0.3$.
Dots denote gauginos ($P > 0.9$), circles denote higgsinos 
($P < 0.1$) and crosses denote mixed ($0.1 \leq P \leq 0.9$)
configurations ($P$ being defined as $P \equiv a_1^2 + a_2^2$).  
}

\end{figure}


\begin{thebibliography}{10}

\bibitem{bf} For a review of neutralino as a dark matter particle,
  see, for instance, 
 A. Bottino and N. Fornengo, {\it Proceedings of the 
 Fifth School on Non-Accelerator Particle Astrophysics} (Abdus
 Salam International Centre for Theoretical Physics, Trieste), 
 Eds. R.A. Carrigan, Jr., G. Giacomelli and N. Paver, E.U.T. 1999, 
[arXiv:hep-ph/9904469]. 



\bibitem{bdfs} A. Bottino, F. Donato, N. Fornengo and S. Scopel, 
{\it Astroparticle Physics} {\bf 13} (2000) 215,
[arXiv:hep-ph/9909228].  


\bibitem{don}
A. Bottino, V. de Alfaro, N. Fornengo, A. Morales, J. Puimedon, S. Scopel, {\it
  Mod.Phys.Lett.} {\bf A7} (1992) 733.


\bibitem{jkg}
G. Jungman, M. Kamionkowski, K. Griest, {\it Phys. Rep.} {\bf 267} (1996) 195.

\bibitem{ellis} J. Ellis, A. Ferstl and K.A. Olive, {\it Phys. Lett.}
  {\bf B481} (2000) 304, [arXiv:hep-ph/0001005].

\bibitem{arnowitt} E. Accomando, R. Arnowitt, B. Dutta and Y. Santoso, 
{\it Nucl.Phys.} {\bf B585} (2000) 124, [arXiv:hep-ph/0001019].

\bibitem{nath} A. Corsetti and P. Nath,  [arXiv:hep-ph/0003186].

\bibitem{fmw} J.L. Feng, K.T. Matchev and F. Wilczek,
  {\it Phys.Lett.} {\bf B482} (2000) 388-399, [arXiv:hep-ph/0004043].

\bibitem{pavan} M.M. Pavan, R.A. Arndt, I.I. Strakovsky and
  R.L. Workman, [arXiv:hep-ph/0111066]. 


\bibitem{qcd_corr} M. Carena, S. Mrenna and E. E. M. Wagner, {\it Phys. Rev.}
  {\bf D60} (1999) 075010; H. Eberl, K. Hidaka, S. Kraml, W. Majerotto and Y.
  Yamada, {\it Phys. Rev.} {\bf D62} (2000) 055006; A. Djouadi and M. Drees,
  {\it Phys. Lett.} {\bf B 484} (2000) 183.


\bibitem{lep} I.M. Fisk and K. Nagai, talks
at the XXXth Int. Conf. on High Energy
Physics, Osaka, July 2000, 
http://www.ichep2000.hep.sci.osaka-u.ac.jp; 
P.J. Donan (ALEPH Collaboration), March 2000, \\
http://alephwww.cern.ch/ALPUB/seminar/lepc\_mar2000/lepc2000.pdf.

\bibitem{cdf} J.A.Valls (CDF Coll.) FERMILAB-Conf-99/263-E CDF; \\ 
http://fnalpubs.fnal.gov/archive/1999/conf/Conf-99-263-E.html. 

\bibitem{barbieri}
R. Barbieri, M. Frigeni and G.F. Giudice, {\it Nucl. Phys.} {\bf B313} (1989)
  725.

\bibitem{griest}
K. Griest, {\it Phys. Rev.} {\bf D38} (1988) 2357, {\it Nucl. Phys.} {\bf B313}
  (1989) 725.

\bibitem{noi1234}
A. Bottino, F. Donato, N. Fornengo and S. Scopel, 
  {\it Phys. Rev} {\bf D59}
  (1999)095003 [arXiv:hep-ph/9808456]. 

\bibitem{svz}
M.A. Shifman, A.I. Vainshstein and V.I. Zacharov, 
{\it Phys. Lett.} {\bf B78} (1978) 443; 
{\it JEPT Lett.} {\bf 22} (1975) 55.

\bibitem{koch1} R. Koch, {\it Z. Phys.}{\bf C15} (1982) 161.

\bibitem{gls2}
J. Gasser, H. Leutwyler and M.E. Sainio, {\it Phys. Lett.} {\bf B253}
(1991) 260.

\bibitem{gl}
J. Gasser, H. Leutwyler, {\it Phys. Rep.}{\bf 87} (1982) 77.

\bibitem{said} SAID pion--nucleon database, http://gwdac.phys.gwu.edu/.

\bibitem{olsson} M.G. Olsson, [arXiv:hep-ph/0001203].

\bibitem{GGR} G.B. Gelmini, P. Gondolo and E. Roulet,
              {\it Nucl. Phys.} {\bf B351} (1991) 623.

\bibitem{probing} 
A. Bottino, F. Donato, N. Fornengo and S. Scopel, 
  {\it Phys. Rev} {\bf D63}
  (2001) 125003 [arXiv:hep-ph/0010203]. 

 
\bibitem{lat} A. Bottino, F. Donato,  N. Fornengo and S. Scopel,
  [arXiv:hep-ph/0105233], to appear in the Proceedings of the 
15th Rencontres de Physique de la Vallee d'Aoste: 
{\it Results and Perspective in Particle Physics}, La Thuile, March 2001.


\bibitem{ge} E. Garcia et al., Phys. Rev. {\bf D51} (1995) 1458;
R. Bernabei et al., Phys. Lett. {\bf B389} (1996) 757, 
 Phys. Lett.  {\bf B 480} (2000) 23, Eur. Phys. J. {\bf C 18} (2000) 283;
Phys. Lett. {\bf B 424} (1998) 195, Phys. Lett. {\bf B436} (1998) 379; 
L. Baudis et al., Phys. Rev. {\bf D59} (1999) 022001;
A. Morales et al., Phys. Lett. {\bf B489} (2000) 268
and [arXiv:hep-ex/0101037]; 
N.J.C. Spooner et al., Phys. Lett. {\bf B473} (2000) 330;  
R. Abusaidi {\it et al.},  {\it Phys. Rev. Lett.} {\bf 84} (2000) 5699; 
A. Benoit et al., [arXiv:astro-ph/0106094].
\bibitem{belli} P. Belli, R. Bernabei, A. Bottino, F. Donato,
  N. Fornengo, D. Prosperi and S. Scopel, 
{\it Phys. Rev.} {\bf D61} (1999) 023512, and references quoted therein.

\end{thebibliography}
\end{document}